\documentclass[12pt]{article}

\usepackage{cite}

\title{
\begin{flushright}
{\normalsize Yaroslavl State University\\
             Preprint YARU-HE-00/04\\
             hep-ph/0009045} \\[10mm]
\end{flushright}
Neutrino-electron processes in a dense maqnetized plasma}

\author{N.V.~Mikheev
\thanks{E-mail address:mikheev@yars.free.net},
E.N.~Narynskaya
\thanks{E-mail address:rose@uniyar.ac.ru},
\\
{\small\it Division of Theoretical Physics, Department of Physics,}\\
{\small\it Yaroslavl State University, Sovietskaya 14,}\\
{\small\it 150000 Yaroslavl, Russian Federation.}}

\date{}

\begin{document}

\maketitle

\begin{abstract}

The neutrino-electron processes in a dense strongly degenerate
magnetized plasma are analyzed in the framework 
of the Standard Model. The total probability and the mean values 
of the neutrino energy and momentum losses are calculated. 
It is shown that neutrino scattering on the excited
electrons  with Landau level number conserva\-ti\-on dominates
under the conditions $ \mu^2 > eB \gg \mu T$ but
does not give a contri\-bu\-tion into the neutrino force acting 
on plasma along the magnetic field.

\end{abstract}

\vglue 5mm

\begin{center}
{\it To appear in Modern Physics Letters A} 
\end{center}

\newpage

It is known that the neutrino physics plays an important role in
astro\-phy\-si\-cal cataclysms like a supernova explosion, a  coalescence
of neutron stars and in the early Universe ~\cite{Raf}.
By this means,  the neutrino processes and the 
neutrino-electron processes in extremal active medium in particular,
are the subject of a great interest~\cite{Coop, Myr, Bez}.
On the other hand, the investigation of the neutrino-electron processes
in such conditions has a conceptual value by itself.

The total set of neutrino-electron processes 
$\nu e^\pm \leftrightarrow \nu e^\pm$,  
$ \nu e^- e^+ \leftrightarrow \nu$,
$ e^\pm \leftrightarrow \nu \bar \nu e^\pm$
in very strong magnetic field when electrons and positrons
occupied the lowest  Landau level, was investigated in the
papers~\cite{Kuz, Mik}. The volume density of the neutrino force
acting on plasma along the magnetic field was calculated in particular.
It was shown that this force was of the same order as the one caused 
by the $\beta $-processes.

In this paper we investigate the  neutrino-electron processes 
in dense strongly degenerate magnetized ultrarelativistic plasma.
In contrast to~\cite{Kuz, Mik}  we consider the
physical situation when the magnetic field is not so strong, 
therefore,  electrons and positrons can occupy  the excited Landau levels.
In this case  the chemical potential of the electrons, $\mu$,
is the largest physical parameter. At the same time it is assumed that the
magnetic field strength is a big parameter also
\footnote{ We use natural units in which $c=\hbar=1$, $e > 0$ is the 
elementary charge. }
\begin{equation}
\mu^2 > eB \gg T^2,E^2 \gg m_e^2.  \label{eq:cond1} 
\end{equation}
Here $T$ is the temperature of plasma,
 $E$ is a typical energy of a neutrino.
It should be noted that such physical conditions could
exist,  for example, in the envelope of an exploding supernova.

Under the conditions (\ref{eq:cond1}) the most part of the
 neutrino-electron processes  is suppressed.
The processes with the electron-positron pair in the 
inital or in the  final states,
$ \nu \bar \nu \leftrightarrow e^- e^+$,
$ \nu  \leftrightarrow \nu e^- e^+$, 
and  the processes where the positron 
presents both in the initial  and in the final state,
$  e^+ \nu \to \nu e^+$,
$  e^+ \leftrightarrow e^+ \bar \nu \nu$, 
are suppressed by statistical factors in the strongly degenerate
plasma.  As for the processes with the electron  both in the initial and 
in the final states, one can show that  the processes with 
the conservation of the Landau level number  are only possible
if the magnetic field strength is large enough.
Really, the conservation  of the
energy, $\varepsilon_{n'}' - \varepsilon_n = q_0$, 
 and of the $z$ component of the  momentum,
 $p'_z - p_z = q_z$, leads to the following condition for the difference
between the Landau level numbers of the initial and the final electrons:
\begin{displaymath}
  n-n' = \frac{q^2_3 - q^2_0 + 2p_zq_z - 2q_0\varepsilon_n}{2eB},
\end{displaymath}
where $\varepsilon_n \simeq \sqrt{p_z^2 + 2eBn}$, $n$  is the
Landau level number, $q_{\alpha}$ is the four-momentum transferred 
from the neutrino to the electron.
The electron energy $\varepsilon_n$ and the component of
 the momenta $p_z$  are of the order of $\mu$ 
and  $q_0$, $q_3$  are of the  order of the  neutrino temperature
$T_\nu \sim T \ll \sqrt{eB}$. Consequently, we can estimate:
\begin{displaymath}
  n-n'  \sim    O   \left( \frac{T\mu}{eB} \right).
\end{displaymath}
If the magnetic field  strength being relatively weak, $eB < \mu^2$, is
simultaneusly strong enough
\begin{equation}
 eB \gg \mu T,  \label{eq:cond2} 
\end{equation}
then  $n = n'$. By this means, the processes of the synchrotron radiation
and absorption of a neutrino pair are forbidden by the energy and
momentum con\-ser\-va\-tion.
The processes of the neutrino-electron scattering only when  both
initial and final electrons occupy the same Landau level
 can be possible under the conditions (\ref{eq:cond1}),(\ref{eq:cond2}).

Numerical calculation
of the differential "cross-section" of the neutrino-electron scattering
in the weak field limit when the magnetic field influence was
unsignificant,  was performed in the paper~\cite{Bez}.
Below we calculate the proba\-bi\-li\-ty of this processes
and the mean values of the neutrino energy and momentum losses.
These calculation could be of interest in a
detalied theoretical description of such astrophysical processes as
a supernove explosion and a coalescence of neutron stars.

If the momentum transferred is relatively small, $|q^2| \ll m^2_W$,
%\footnote{As the analysis shows, it corresponds in this case to the 
and the magnetic field is not so strong, 
$ eB \ll m^2_W $, $m^3_W/T$,
then  weak interaction of neutrino with electrons 
in the framework of SM can be described in the local 
limit by the effective Lagrangian of the following form
\begin{equation}
  L_{eff}= \frac{G_F}{\sqrt2}
  [\overline e \gamma_{\alpha}(g_v - g_a\gamma_5)e]
  [\overline \nu\gamma_{\alpha}(1 - \gamma_5)\nu],
\label{eq:lag}
\end{equation}
where $g_v=\pm1/2 + 2sin^2\theta_W$, $g_A=\pm 1/2$. Here upper signs correspond
to the electron neutrino $(\nu = \nu_e)$ when both Z and W boson exchange takes
part in a process. The lower signs correspond to $\mu$ and $\tau$
neutrino $(\nu = \nu_\mu,\nu_\tau)$, when the Z bozon exchange is only presented
in the Lagrangian (\ref{eq:lag}).

The amplutide of the neutrino-electron scattering process can be obtained 
immediately from the  Lagrangian (\ref{eq:lag}) 
 by means of substitution of well known
 solutions of the Dirac equation in a magnetic field,
 see, for example~\cite{Ax}.

After the summation over the spin states of both the initial and 
the final electrons which occupy the same excited
Landau levels  with number $n$,
the S-matrix element squared is:
\begin{equation}
\sum\mid S \mid^2 \simeq  32G_F^2(g_v^2+g_a^2)\pi^3 {\cal T} eBn    
 \frac{[2(k\tilde\varphi q)^2+ q^2_{\parallel}(kk')]}
    {E E' \varepsilon_n \varepsilon'_n L_yL_zV^2(-q^2_{\parallel})}
\delta^3().
\end{equation}
Here 
${\cal T}$ is the total interaction time,
$\delta^3()=\delta(\varepsilon_n-\varepsilon'_n+q_0)\delta(p_z-p_z'+q_z)
\delta(p_y-p'_y+q_y)$ corresponds to the energy and momentum 
 conservation law in the presence of an external
 magnetic field directed along the $z$  axis
in the gauge $A^\mu = (0,0,Bx,0)$,
$k^{\alpha}=(E,\vec k)$, $k'^{\alpha}=(E',\vec k')$  are 
the 4-vectors of energy-momentum of the initial and the final
 neutrino correspondingly, 
$q^{\alpha}=k^{\alpha} - k'^{\alpha}$,
$q^2_{\parallel}=q \tilde\Lambda q = q_0^2 - q_3^2$,
$\varphi_{\alpha\beta}=F_{\alpha\beta}/B$ and
$\tilde\varphi_{\alpha\beta}=\epsilon_{\alpha\beta\mu\nu}\varphi_{\mu\nu}/2$
are the dimensionless  magnetic field tensor and the dual tensor,
$\tilde\Lambda_{\alpha\beta}=
(\tilde\varphi \tilde\varphi)_{\alpha\beta}=
\tilde\varphi_{\alpha\gamma}\tilde\varphi_{\gamma\beta}$,
 $V=L_xL_yL_z$   is the normalization volume.

The probability  of the neutrino scattering on 
plasma electrons  per unit time has a physical meaning only being integrated
over both the final and the initial electron states as well
\begin{equation}
W(\nu e^- \to \nu e^-)= 
\frac{1}{{\cal T}} \int \mid S\mid^2 
  d\Gamma_{e^-}  f_{e^-} d\Gamma'_{e^-}(1-  f'_{e^-}) d\Gamma'_{\nu} (1-  f'_{\nu}),
  \label{eq:ver1}
\end{equation}
here $d\Gamma$ is the phase-space
element of a particle, $f$ is its distribution function,
$  f_{\nu} = [exp(E - \mu_\nu)/T_\nu) + 1)]^{-1}$,
$\mu_\nu$ and $T_\nu$ are the chemical potential and the temperature of the neutrino
gas correspondingly, $f_{e^-} = [exp( \varepsilon - \mu)/T) + 1)]^{-1}$.
In a general case the neutrino temperature $T_\nu$ can differ from
 the plasma temperature $T$.
                             
The integration over the momentum space of all electrons can be performed by the
following way. The integration over the $z$ and $y$ components of the final electron removes the 
momentum $\delta$-functions. Then we integrate over the $z$ component of the initial 
electron momentum taking into account
 the energy $\delta$-function:
\begin{displaymath}
 \frac{\delta(\varepsilon_n-\varepsilon_n'+q_0)}{\varepsilon_n\varepsilon'_n}
   \simeq \frac{\delta(p_z-p^*)}{\sqrt{2eBn}\sqrt{-q^2_{\parallel}}},
   \qquad
   p^* \simeq \frac{q_3q_0\sqrt{2eBn}}{\mid q_3 \mid \sqrt{-q^2_{\parallel}}}.
\end{displaymath}

The integration over the $y$ component of the initial electron momentum
 can be performed due to the fact that the integrand does not
depent on $p_y$ so this integration gives
in fact a power of a  degeneration, $s$, of the state with the energy 
 $\varepsilon_n$:
\begin{displaymath}
 s = 2  \int \frac{L_ydp_y}{2\pi} = \frac{eBL_xL_y}{\pi}.
\end{displaymath}
In view of the integration over the final neutrino momentum
it is useful to note that:

 i) the electron distribution function 
    in the  strongly degenerate plasma, when $\mu \gg T$,
    has a form which is close to the stepwise;

 ii) the electrons from a narrow region near the Fermi level only
 take part in the scattering processes if the
 neutrino energy is relatively small, $E \sim T_\nu \ll \mu$.

Therefore, 
the product of the statistical factors of the initial and the final 
electrons can be extrapolated in this case by the $\delta$-function:
\begin{displaymath}
 f(\varepsilon)(1-f(\varepsilon')) 
  \simeq \frac{q_0}{(1-e^{-q_0/T})}\delta(\varepsilon -\mu).
\end{displaymath}
With the energy conservation, this product can be written in the form
\begin{displaymath}
 f(\varepsilon)(1-f(\varepsilon')) 
  \simeq \frac{q_0\mid q_0 \mid}{(1-e^{-q_0/T})}
  \frac{(1-z^2)}{\mu E' z^3}
 [\delta(cos\theta'-c_1)+\delta(cos\theta'-c_2)].
\end{displaymath}
Here 
$z=\sqrt{1-2eBn/\mu^2}$,$c_{1,2}=(k_3' \pm q_0/z)/E'$,
$\theta'$ is the angle between the final neutrino
 momentum $\vec k'$ and the  magnetic field direction.

 The total probability of the  neutrino-electron processes is
 the sum over the Landau levels:
\begin{displaymath}
    W_{tot} = W_{n=0} +  W_{n \ge 1}. 
\end{displaymath}
  Here $W_{n=0}$ is the probability  of the neutrino-electron scattering 
when electrons  occupy only the ground Landau
level~\cite{Kuz, Mik}, $ W_{n \ge 1} = \sum \limits_{n=1}^{n_{max}}W_{n \to n}$
is the contru\-bu\-tion of the excited Landau levels, 
$n_{max}$ is the integer part of the ratio $ \mu^2 /2eB \ge 1 $.
The result of our calculation for the probability  $W_{n \ge 1}$ is:
\begin{eqnarray}
  W_{n \ge 1} = \frac{G_F^2(g_v^2+g_a^2) eB T^2E}{4\pi^3}
  & \sum  \limits_{n=1}^{n_{max}} & \frac{1}{z^2} \bigg \lbrace  
  ((1+z^2)(1+u^2)-4uz)\int \limits_{a}^{b} \Phi (\xi) d\xi 
\label{eq:prob}
\nonumber\\
 - \frac{1}{zr \tau}(1-z^2)(z-u) & \int \limits_{a}^{b} & \xi 
   \Phi (\xi) d\xi  \,\,\,  + \,\,\, u \to -u  \bigg \rbrace,
\end{eqnarray}
where
$\Phi (\xi)=\xi[(e^\xi-1)(e^{\eta_\nu-r-\xi/\tau}+1)]^{-1}$,
$a=-r\tau z(1+u)/(1+z)$ and  $b=r\tau z(1-u)/(1-z)$,
$ r=E/T_\nu$, $\tau =T_\nu/T$, $\eta_\nu=\mu_\nu/T_\nu$,
$u=cos\theta$, $\theta$ is the angle between the initial neutrino
 momentum $\vec k$ and the magnetic field direction.

In the limit of  a rare neutrino gas, $E \gg T_\nu$,
and very dense plasma, $\mu^2 \gg eB$, when the electrons
of the highest Landau levels give the main contribution into the
probability, the result has a simple form:
\begin{equation}
W_{tot} \simeq \frac{G^2_F\mu^2(g_v^2+g_a^2)E^3}{15 \pi^3}
          (0,82 u^4 - 1,71 u^2 + 1,11).
\end{equation}
One has to remember that  the magnetic field is not-to-small, $eB \gg \mu E$.
Despite the fact that this expression does not depend on the
value of the magnetic field strength,
the probability  is not isotropic and
 does not coincide with the result without field:
\begin{equation}
W_{vac}=\frac{G^2_F\mu^2(g_v^2+g_a^2)E^3}{15\pi^3}.
\end{equation}

It should be noted that a practical significance
for astrophysics can be in the mean values of the density of
 neutrino energy  
and the momentum losses  per unit time
in a medium rather than in the process probabilities.
These mean values could be  defined as
\begin{equation}
  (\dot \varepsilon, {\cal F}) = 
       \frac{1}{(2\pi)^3}\int \frac{q_{0,3}d^3k}{(e^{(E-\mu_\nu)/T_\nu}+1)}\,dW ,
\end{equation}
where $\dot \varepsilon$ determines the energy transferred from 
a neutrino to the 
plasma in unit volume per unit time,
${\cal F}$ defines the loss of the neutrino momentum
 in unit volume per unit time
and therefore it defines the neutrino force 
acting on plasma along the magnetic field.

For the  neutrino  energy loss in unit volume per unit time
we obtain the following result:
\begin{eqnarray}
 \dot \varepsilon_{n \ge 1}  & = & \sum \limits_{n=1}^{n_{max}}
   \dot \varepsilon_{n \to n}    = 
 \frac{G_F^2(g_v^2+g_a^2) eB T^7_\nu}{12\pi^5} 
 \sum \limits_{n=1}^{n_{max}} \frac{1}{z^3} 
   \int \limits_{0}^{\infty}  \xi^2 d\xi 
\frac{ e^{\xi \tau} - e^{\xi}}{e^{\xi \tau } - 1}
 \label{eq:eps} \\
& \times  & \int \limits_{0}^{\infty}  
  \frac{y^2 (3\xi (1-z^2) + 4zy(1+z^2)) dy}
   {(e^{\eta_\nu - y - \xi (1-z)/2z} + 1)
    (e^{ - \eta_\nu + y + \xi (1+z)/2z} + 1)} 
\nonumber
\end{eqnarray}
The formula (\ref{eq:eps}) demonstrates that in a case of an equilibrium
of the neurtrino gas with electron plasma, $\tau=T_\nu/T \to 1$, the neutrino energy loss 
per unit time tends to zero.

The probability of the neutrino-electron scattering 
$W_{n \ge 1}$ (\ref{eq:prob})
is symmetric with respect to the substitution $\theta \to \pi - \theta$.
It means that the neutrino scattering on electrons which occupy the 
excited Landau levels does not give a contribution into the force ${\cal F}$.
 Therefore,
this force is caused by a contribution of neutrino  scattering  on the ground 
Landau level
electrons only, and the result for this force obtained
in~\cite{Mik} has a more general applicability in fact.
 It may be used even though $\mu^2$ is considerably greater than the magnetic
field strength.

\bigskip

We are grateful to A.~Kuznetsov for helpful discussions.

This work was supported in part by Russian Foundation for Basic Research 
Grant N~98-02-16694.

\end{document}